\documentclass[aps,twocolumn,superscriptaddress,longbibliography,nofootinbib]{revtex4-2}

\usepackage{amsmath}
\usepackage{amssymb}
\usepackage{amstext}
\usepackage{amsopn}
\usepackage{amsfonts}
\usepackage{amsxtra}
\usepackage[english]{babel}
\usepackage{graphicx}
\usepackage{bm}
\usepackage{multirow}
\usepackage{dcolumn}
\usepackage{color}
\usepackage{verbatim}
\begin{document}

\title{Charge Density Waves in the 2.5-Dimensional Quantum Heterostructure}

\author{F. Z. Yang}
\affiliation{Materials Science and Technology Division, Oak Ridge National Laboratory, Oak Ridge, Tennessee 37831, USA}
\author{T. T. Zhang}
\affiliation{Institute of Theoretical Physics, Chinese Academy of Sciences, Beijing 100190, China}
\author{F. Y. Meng}
\affiliation{Department of Physics and Beijing Key Laboratory of Opto-electronic Functional Materials $\&$ Micro-nano Devices, Renmin University of China, Beijing 100872, China}
\affiliation{Key Laboratory of Quantum State Construction and Manipulation (Ministry of Education), Renmin University of China, Beijing 100872, China}
\author{H. C. Lei}
\affiliation{Department of Physics and Beijing Key Laboratory of Opto-electronic Functional Materials $\&$ Micro-nano Devices, Renmin University of China, Beijing 100872, China}
\affiliation{Key Laboratory of Quantum State Construction and Manipulation (Ministry of Education), Renmin University of China, Beijing 100872, China}
\author{C. Nelson}
\affiliation{National Synchrotron Light Source II, Brookhaven National Laboratory, Upton, New York 11973, USA}
\author{Y. Q. Cai}
\affiliation{National Synchrotron Light Source II, Brookhaven National Laboratory, Upton, New York 11973, USA}
\author{E. Vescovo}
\affiliation{National Synchrotron Light Source II, Brookhaven National Laboratory, Upton, New York 11973, USA}
\author{A. H. Said}
\affiliation{Advanced Photon Source, Argonne National Laboratory, Lemont, Illinois 60439, USA}
\author{P Mercado Lozano}
\affiliation{Advanced Photon Source, Argonne National Laboratory, Lemont, Illinois 60439, USA}
\author{G. Fabbris}
\affiliation{Advanced Photon Source, Argonne National Laboratory, Lemont, Illinois 60439, USA}
\author{H. Miao}\email[]{miaoh@ornl.gov}
\affiliation{Materials Science and Technology Division, Oak Ridge National Laboratory, Oak Ridge, Tennessee 37831, USA}

\date{\today}


\begin{abstract}

Charge density wave (CDW) and their interplay with correlated and topological quantum states are forefront of condensed matter research. The 4$H_{b}$-TaS$_2$ is a CDW ordered quantum heterostructure that is formed by alternative stacking of Mott insulating 1T-TaS$_2$ and Ising superconducting 1H-TaS$_2$. While the $\sqrt{13}\times\sqrt{13}$ and 3$\times$3 CDWs have been respectively observed in the bulk 1T-TaS$_2$ and 2H-TaS$_2$, the CDWs and their pivotal role for unconventional superconductivity in the 4$H_{b}$-TaS$_2$ remain unsolved. In this letter, we reveal the 2-dimensional (2D) $\sqrt{13}\times\sqrt{13}$ chiral CDW in the 1T-layers and intra-unit cell coupled 2D 2$\times$2 CDW in the 1H and 1H' layers of 4$H_{b}$-TaS$_2$. Our results establish 4$H_{b}$-TaS$_2$ a novel 2.5D quantum heterostructure, where 2D quantum states emerge from 3D crystalline structure.


\end{abstract}

\maketitle

Charge density wave (CDW), an electronic liquid that spontaneously breaks translational symmetry of the underlying lattice, plays a key role in correlated quantum materials, including cuprate high-T$_c$ superconductors \cite{Tranquada1995, Ghiringhelli2012, Miao2017} and topological semimetals \cite{Wang2013, Gooth2019, Tang2019}. Recently, superconductivity that possibly breaks the time-reversal symmetry, $\mathcal{T}$, and translational symmetry, $\mathcal{R}$, is observed in a CDW ordered quantum heterostructure 4$H_{b}$-TaS$_2$ \cite{Ribak2020, Nayak2021,Persky2022,Luo2024, Yang2024}, raising questions on the interplay between intertwined symmetry-breaking orders and the stacking degree of freedom. 

\begin{figure}
\includegraphics[width=1\linewidth]{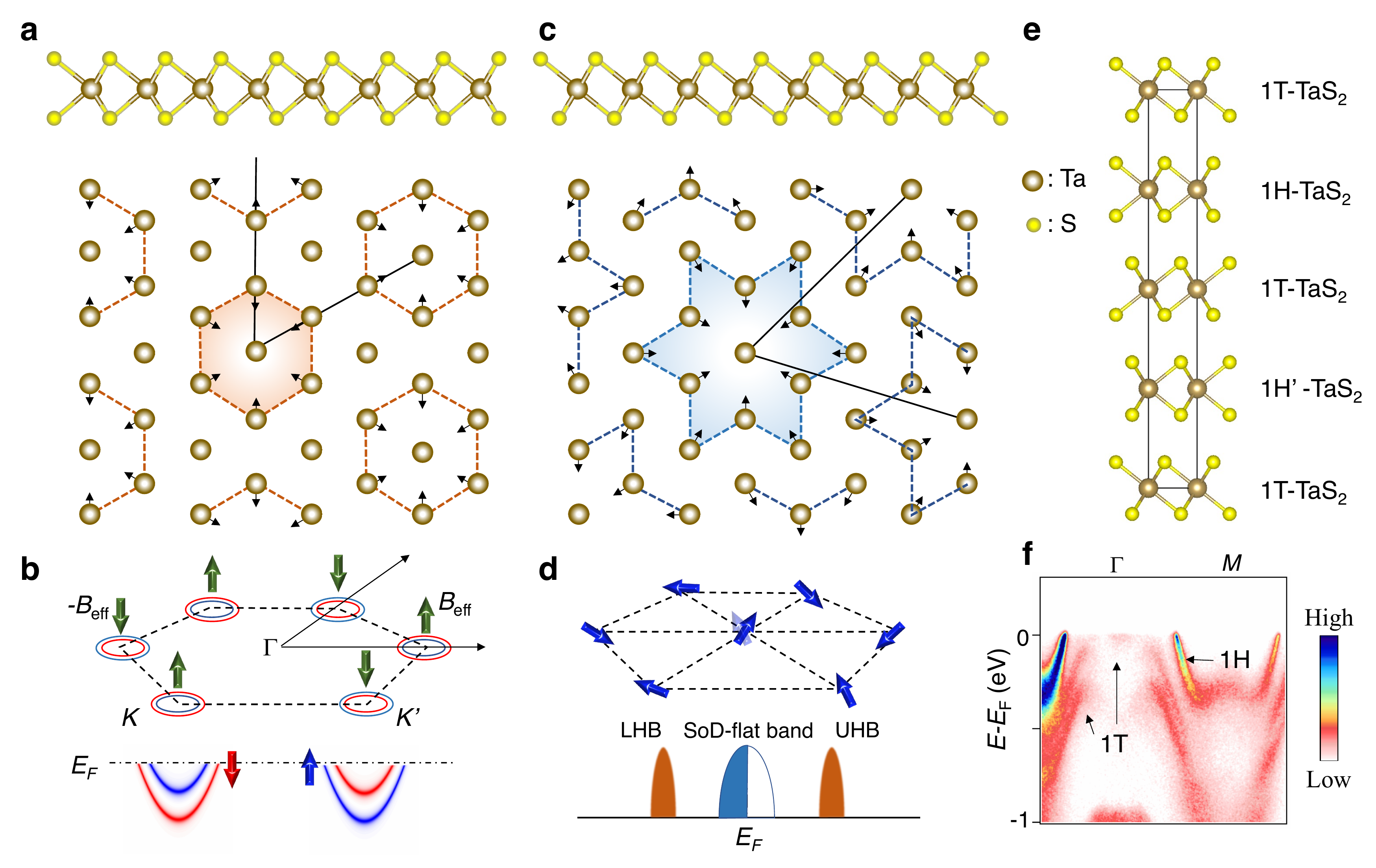}
\caption{(a) Side view of monolayer 1H-TaS$_2$ that breaks $\mathcal{P}$. Top view of Ta triangular lattice that displays 3$\times$3 CDW \cite{BROUWER1980, Scholz1982}. (b) The electronic structure of 1H-TaS$_2$. The Ising-type SOC induces effective local magnetic field, $B_{eff}$, in the momentum space that lifts the spin degeneracy. (c) Side view of monolayer 1T-TaS$_2$ that preserves $\mathcal{P}$. (d) The SoD CDW induced band folding results in single-occupied flat band that eventually forms the upper Hubbard band (UHB) and lower Hubbard band (LHB). (e) Crystal structure of 4$H_{b}$-TaS$_2$. (f) ARPES determined electronic structure confirms the coexistence of 1H and 1T electronic states.}
\label{fig1}
\end{figure}

Structurally, the 4$H_{b}$-TaS$_2$ is formed by periodic stacking of 1H and 1T-TaS$_2$ layers. Figure~\ref{fig1}a shows the crystal structure and 3$\times$3 CDW of the monolayer 1H-TaS$_2$. Due to the absence of inversion symmetry, $\mathcal{P}$, the Ising-type spin-orbital coupling (SOC) lifts the spin degeneracy near the K and K' point in the momentum space (Fig.~\ref{fig1}b) \cite{Xiao2007,Xu2014} and gives rise to Ising superconductivity below $T_{SC}\sim$3~K \cite{Lu2015, Xi2016, Li2021, Zhang2022, Barrera2018}. Figure~\ref{fig1}c shows the structure of $\mathcal{P}$-preserving 1T-TaS$_2$ that hosts $\sqrt{13}\times\sqrt{13}$ CDW and star-of-David (SoD) lattice distortions. The CDW induces single-occupied flatband is pushed away from the Fermi level, $E_{F}$, by forming a Mott gap, that possibly hosts Dirac quantum spin liquid (Fig.~\ref{fig1}d) \cite{Law2017, Ruan2021}. By adding the stacking degree of freedom to combine the CDW ordered 1H and 1T-TaS$_2$ layers, the 4$H_{b}$-TaS$_2$ offers a unique opportunity to realize intertwined topological quantum phases \cite{Ribak2020, Nayak2021,Persky2022,Luo2024, Yang2024}. Here, we use hard x-ray diffraction (XRD) to establish 4$H_{b}$-TaS$_2$ as a novel 2.5-dimensional (2.5D) quantum heterostructure, where 2D electronic states and CDWs emerge in the 3D crystal structure \cite{Ago2022}.

\begin{figure*}
\includegraphics[width=1\linewidth]{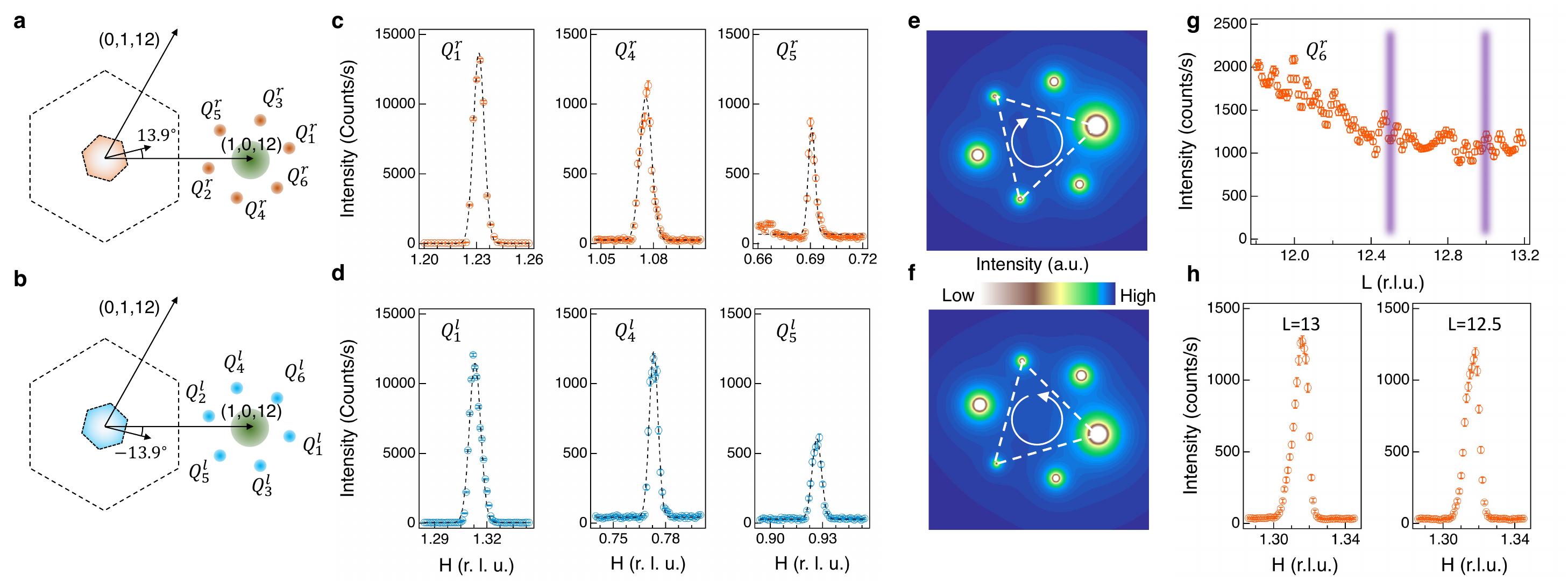}
\caption{(a) and (b) depicts Brillouin zones of undistorted 1T-TaS$_2$ (dashed hexagon), right-handed CDW phase (pink hexagon), and left-handed CDW phase (blue hexagon). $Q^{r/l}=q^{r/l}+\mathbf{G}$ represent the right and left-handed CDW superlattice peak positions near the $\mathbf{G}$=(0, 1, 12) fundamental Bragg peak. (c) and (d) shows the H-scan of CDW superlattice peaks. Dashed lines are fittings of the CDW peak intensity using a Gaussian function. (e) and (f) summarize the right and left-handed CDW peak intensity distributions in the HK scattering plane. Both the peak positions and intensity distributions of the right and left-handedness are related by $M_y$ mirror symmetry. The white triangles connect CDW superlattice peaks, whose intensity distribution reflects the handedness of the CDW domain. (g) and (h) show $L$ and $H$ scans at $Q_{6}^{r}$, establishing 2D CDW in the 1T layers. Error bars represent 1 standard deviation assuming Poisson counting statistics.} 
\label{fig2}
\end{figure*}

High-quality single crystals of 4$H_{b}$-Ta(S, Se)$_2$ were grown using the chemical vapor transport method \cite{Meng2024}. The elastic x-ray diffraction measurements were performed at the beamline 4-ID-D at the Advanced Photon Source and the beamline 4-ID at National Synchrotron Light Source II (NSLS-II). The photon energy was set to $h\nu=$9.88 keV. The inelastic x-ray scattering (IXS) was conducted at beamline 10-ID at the NSLS-II. The highly monochromatic x-ray beam of incident energy $E_{i}=$9.13~keV ($\lambda = 1.36$\AA{}) was focused on the sample with a beam cross-section of $\sim10\times10$~$\mathrm{\mu m}^{2}$. The total energy resolution was $\Delta E\sim 1.4$~meV (full width at half maximum). The measurements were performed in reflection geometry. Typical counting times were in the range of 120 to 240 seconds per point in the energy scans at constant momentum transfer $\textbf{Q}$. The ARPES experiments were performed at beamline 21-ID-1 of NSLS-II. Samples were cleaved $in~situ$ in a vacuum better than 3$\times$10$^{-11}$ Torr. The total energy resolution of ARPES measurement was approximately 15~meV. The sample stage was maintained at 30~K throughout the experiments. The force constants for monolayer 1H-TaS2 were calculated using density functional perturbation theory (DFPT) and the Vienna Ab initio Simulation Package (VASP). The force constant was used to obtain the phonon dispersion and dynamical structure factor. Prior to the force constants calculations, the crystal structure is relaxed with the residual force on each atom less than 0.001 eV/\AA. The exchange correlation potential was treated within the generalized gradient approximation (GGA) of the Perdew-Burke-Ernzerhof variety, where the kinetic energy cutoff was set to 420~eV. Integration for the Brillouin zone was done by using a MonkhorstPack k-point grids which is equivalent to $12\times 12\times 1$ in the self-consistent calculation.

Figure~\ref{fig1}f shows the ARPES determined electronic structure of 4$H_{b}$-TaS$_2$ along the $\Gamma-$M direction, confirming the coexistence of 1H and 1T electronic states \cite{Ribak2020, Almoalem2024}. Due to the charge transfer between 1H and 1T layers, the Mott gap in the 1T layer is collapsed leaving the flat band on the, $E_{F}$. Figure~\ref{fig2} determines the $\sqrt{13}\times\sqrt{13}$ CDW in the 1T layers \cite{BROUWER1980, Scholz1982}. As shown in Fig.~\ref{fig1}e, the SoD lattice distortions break the mirror symmetry of the Ta triangular lattice giving rise to right-handed and left-handed domains as depicted in Fig.~\ref{fig2}a and b. For the right-handed CDW domains, the CDW peaks are located at $q_{1}^{r}$=-$q_{2}^{r}$=(3/13, 1/13), $q_{3}^{r}$=-$q_{4}^{r}$=(-1/13, 4/13), and $q_{5}^{r}$=-$q_{6}^{r}$=(-4/13, 3/13) in the reciprocal lattice unit (r.l.u.). For the left-handed CDW domains, the CDW peaks are located at $q_{1}^{l}$=-$q_{2}^{l}$=(4/13, -1/13), $q_{3}^{l}$=-$q_{4}^{l}$=(3/13, -4/13), and $q_{5}^{l}$=-$q_{6}^{l}$=(-1/13, -3/13) r.l.u. Here $q^{r/l}$ are reduced momentum transfers that are related to $Q^{r/l}$ (Fig.~\ref{fig2}a and b) by the reciprocal lattice unit, $\mathbf{G}=$(0, 1, 12). Figures~\ref{fig2}c and d show the CDW superlattice peaks of the right and left-handed domains. Dashed curves are fittings using a Gaussian function that yield half-width-at-half-maximum (HWHM)$\sim$0.004~r.l.u. and in-plane CDW correlation length, $\xi_{//}=$1/HWHM$\sim$130~\AA. Figures~\ref{fig2}e and f depict the extracted right and left-handed CDW peak intensities in the 2D HK-plane. Both the superlattice peak positions and intensities of the left and right-handed domains are related by the mirror, $M_y$, symmetry, as expected for the scattering near the (0, 1, 12) Bragg peak. Interestingly, we find that the chirality of the CDW domain is encoded in the diffraction intensities. As highlighted by the white dashed triangles in Fig~\ref{fig2}e and f, the peak intensity of $q_{1}^{r}$, $q_{4}^{r}$, and $q_{5}^{r}$ decreases clockwise (Fig.~\ref{fig2}c), whereas the the peak intensity of $q_{1}^{l}$, $q_{4}^{l}$, and $q_{5}^{l}$ decreases counterclockwise (Fig.~\ref{fig2}d). These chiral peak intensities arise from the CDW-induced SoD lattice distortions and can be reproduced by calculating x-ray scattering amplitude \cite{Supp}. Most surprisingly, we find that despite the intermediate CDW correlation length in the HK plane, the CDW is decoupled along the crystalline c-axis as revealed by the L-scan shown in Fig.~\ref{fig2}g and H-scans at L=13 and 12.5 shown in Fig.~\ref{fig2}h. These observations starkly contrast the bulk 1T-TaS$_2$ that forms 3D CDW \cite{BROUWER1980, Burk1992}, and establish a rare 2D CDW in 3D structure.

\begin{figure}
\includegraphics[width=1\linewidth]{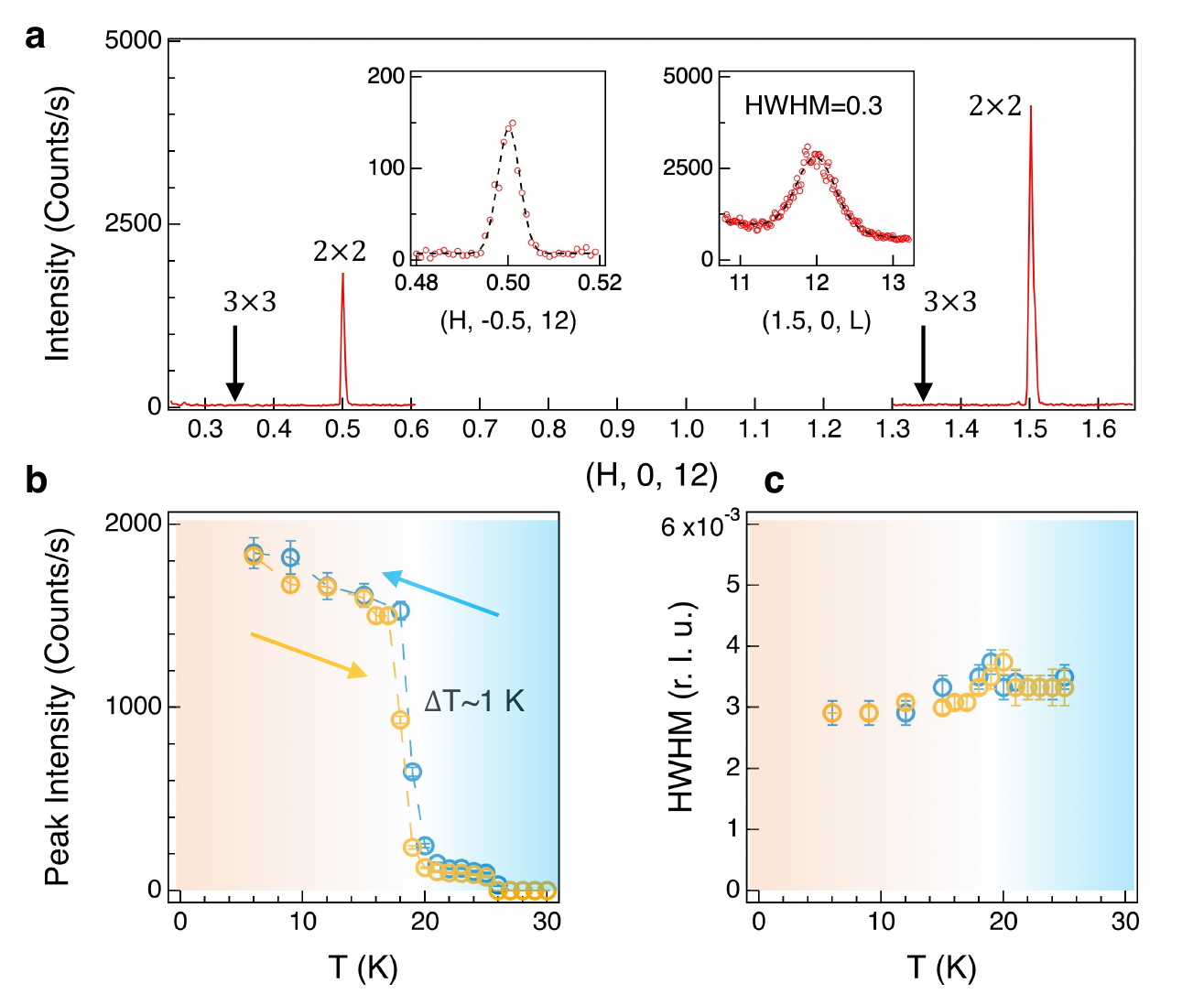}
\caption{(a) XRD along the (H, 0, 12) direction at $T=$10~K shows $2\times2$ CDW. Inset shows the K-scan near the (0.5, -0.5, 12) superlattice peak position and L-scan near the (1.5, 0, 12) superlattice peak position. Dashed lines are Gaussian function fittings of the experimental data. The extracted in-plane and out-plane correlation lengths are 170 and 13~\AA, respectively. (b) and (c) show the temperature-dependent peak intensity and HWHM, respectively. Yellow and cyan data were collected by warming and cooling respectively. Since the $2\times2$ CDW is broadly peaked at $L=even$~$integers$, the intra-unit cell coupled CDW in the 1H and 1H' layers are in-phase. The 1~K temperature hysteresis supports a weak first-order transition.} 
\label{fig3}
\end{figure}

We now focus on CDWs in the 1H layers. Figure~\ref{fig3}a shows the XRD intensity along the (H, 0, 12) direction. Previous scanning tunneling microscopy studies of 4$H_{b}$-TaS$_2$ show $3\times3$ CDW on the surface 1H-TaS$_2$ \cite{Nayak2021, Shen2022}, in agreement with monolayer 1H-TaS$_2$ and bulk 2H-TaS$_2$\cite{BROUWER1980, Scholz1982}. Unexpectedly, our bulk sensitive XRD shows CDW superlattice peaks at H=$Integer+$0.5, corresponding to $2\times2$ CDW in the 1H layers of 4$H_{b}$-TaS$_2$. The insets of Fig.~\ref{fig3}a show x-ray diffraction intensity near (0.5, -0.5, 12) and along the (1.5, 0, L) direction. The extracted in-plane and out-plane correlation lengths are $\xi_{//}\sim$170~\AA~and $\xi_{\perp}\sim$13~\AA, respectively. While the $2\times2$ CDW in the 1H layers displays a peak at $L$=12, the short $\xi_{\perp}$ reveals an in-phase and intra-unit cell coupled $2\times2$ CDW between the 1H and 1H' layers. Figures~\ref{fig3}b and c show the temperature dependence of CDW peak intensity and peak width, respectively. In agreement with resistivity measurement \cite{Supp}, we find that the CDW peak intensity displays a sudden jump at $T_{1H}$=20~K. Warming (yellow) and cooling (cyan) of the sample reveal small hysteresis, $\Delta T\sim$1K, confirming a weak first-order phase transition \cite{LiH2021, Miao2021_2, Li2022, Miao2023}. Interestingly, despite the significantly suppressed peak intensity above $T_{1H}$=20~K, the CDW peak remains observable until $T^{*}=$25~K. We attribute the persistent CDW to competing CDW instabilities in the 1H layers \cite{Supp}.

Our observations of 2D chiral CDW in the 1T layers and intra-unit cell coupled 2D $2\times2$ CDW in the 1H and 1H' layers make 4$H_{b}$-TaS$_2$ as a novel 2.5-dimensional (2.5D) quantum heterostructure \cite{Ago2022}. The emergence of the 2D electronic states in the 3D bulk structure has significant implications for low-temperature quantum states. The $\sqrt{13}\times\sqrt{13}$ chiral CDW in the 1T-layers yields 2D chiral flat band that is pinned at the $E_F$ \cite{Yang2022, Almoalem2024}. Combining with the proximity induced superconductivity, $\mathcal{T}$-breaking superconducting pairing may emerge in the 1T-layers \cite{Ribak2020, Nayak2021,Persky2022,Luo2024}. On the other hand, the intra-unit cell coupled CDW forms an effective 2D bilayer structure, which has been predicted to host orbital Fulde-Ferrell-Larkin-Ovchinnikov state \cite{LiuCX2017, Wan2023, Xie2023}. Furthermore, the formation of intra-unit cell 1H-1H' bilayer breaks the global mirror, $M_z$, symmetry of 4$H_{b}$-TaS$_2$ allowing the Lifshitz invariant term in the Ginzburg-Landau free energy \cite{Edelstein1996}. This Lifshitz invariant can stabilize finite momentum pairing state in 3D systems under external magnetic field \cite{Yang2024}.

In summary, we revealed 2D chiral CDW in the 1T-layers and intra-unit cell coupled 2D CDW in the 1H layers. Our results highlight 4$H_{b}$-TaS$_2$ as a 2.5D quantum heterostructure that can host emergent macroscopic quantum phases. 

Acknowledgments: We thank Gang Chen, Hong Ding, Patrick A. Lee, Chao-Xing Liu, Lingyuan Kong, Wenyao Liu, Qiangsheng Lu, Ziqiang Wang, Binghai Yan, Jiaqiang Yan, Noah F. Q. Yuan, Rui-Xing Zhang, and Yang Zhang for stimulating discussions. This research was supported by the U.S. Department of Energy, Office of Science, Basic Energy Sciences, Materials Sciences and Engineering Division (X-ray and ARPES). X-ray scattering used resources (beamline 4-ID and 30-ID) of the Advanced Photon Source, a U.S. DOE Office of Science User Facility operated for the DOE Office of Science by Argonne National Laboratory under Contract No. DE-AC02-06CH11357. ARPES and X-ray scattering measurements used resources at 21-ID-1, 4-ID and 10-ID beamlines of the National Synchrotron Light Source II, a US Department of Energy Office of Science User Facility operated for the DOE Office of Science by Brookhaven National Laboratory under contract no. DE-SC0012704. H.C.L. was supported by Beijing Natural Science Foundation (Grant No. Z200005), National Key R\&D Program of China (Grants No. 2022YFA1403800, No. 2023YFA1406500), National Natural Science Foundation of China (Grants No. 12274459).

\bibliography{ref}
\end{document}